\documentclass[11pt]{article}
\usepackage{geometry}                
\usepackage{graphicx}
\usepackage{amssymb}
\usepackage{epstopdf}
\title{Spin-orbit resonances and rotation of coorbital bodies in quasi-circular orbits}
\author{Philippe Robutel\footnote{IMCCE, Observatoire de Paris, UPMC Univ. Paris 06, Univ. Lille 1, CNRS, 77~Av.Denfert-Rochereau, 75014 Paris, France }, \, 
   Alexandre C.M. Correia\footnote{Departamento de F\'isica, I3N, Universidade de Aveiro, Campus de
Santiago, 3810-193 Aveiro, Portugal }$^{\phantom{*}*}$,\, 
   Adrien Leleu$^*$}

\date{September 10, 2014}

\begin{document}
\maketitle

\begin{abstract}
The rotation of asymmetric bodies in eccentric Keplerian orbits can be chaotic when there is some overlap of spin-orbit resonances. Here we show that the rotation of two coorbital bodies (two planets orbiting a star or two satellites of a planet) can also be chaotic even for quasi-circular orbits around the central body. When dissipation is present, the rotation period of a body on a nearly circular orbit is believed to always end synchronous with the orbital period. Here we demonstrate that for coorbital bodies in quasi-circular orbits, stable non-synchronous rotation is possible for a wide range of mass ratios and body shapes. We further show that the rotation becomes chaotic when the natural rotational libration frequency, due to the axial asymmetry, is of the same order of magnitude as the orbital libration frequency.

\end{abstract}

\section{Rotation on a Keplerian orbit}

 Considering that the spin axis of a body is perpendicular to the orbital plane, the orientation of the long axis in a inertial reference frame is given by the angle $\theta$ satisfying the differential equation (e.g. Goldrich \& Peale, 1966):
\begin{equation}
\ddot{\theta} + \frac{\sigma^2}{2}\left(\frac{a}{r}\right)^3  \sin{2(\theta - f)} =0 , \,\, {\rm with} \,\, \sigma =  n\sqrt{\frac{3(B-A)}{C}} \, ,
\label{eq:rot_gene}
\end{equation}
where $A<B<C$ are the internal momenta of the body, $(r,f)$ the polar coordinates of the center of the body,  $a$ and $n$ its instantaneous semi-major axis and mean motion. 
One of the simplest situation is the well known case of a rotating body orbiting the primary in Keplerian motion. Then,  the equation (\ref{eq:rot_gene}) can be expanded in Fourier series of the time, given the expression: 
\begin{equation}
\ddot \theta + \frac{\sigma^2}{2}\sum_{k = -\infty}^{+\infty} X_k^{-3,2}(e)\sin(2\theta - k nt) = 0 \, ,
\label{eq:rot_kep}
\end{equation}
where the Hansen coefficients $X_k^{-3,2}(e)$ are of order $ e^{\vert k -2\vert}$ as the eccentricity $e$ tends to zero. 
The equation (\ref{eq:rot_kep}) imposes that the main resonances, namely the spin-orbit resonances, are centered at $\dot\theta =  k  n/2 $ where their width, in terms of frequency, is of order $\sigma \sqrt{e}^{\vert k-2\vert}$. As a consequence, the three largest spin-orbit resonances are, at least for small to moderate eccentricity, the 1:1, 3:2 and 1:2.  For small values of $\sigma$  or $e$, the islands are isolated and the global dynamics is regular in the concerned region.  If the width of the resonance islands increases, for example by increasing the axial asymmetry, and consequently $\sigma$, or the eccentricity of the Keplerian orbit, the resonances partially overlap to give rise to a huge chaotic sea. This kind of phenomenon was described  by Wisdom et al. (1984) in the case of Hyperion.  In contrast, if the eccentricity is decreased to zero the width of all spin-orbit resonances, except the 1:1 which corresponds to the synchronous rotation, shrinks to zero.  In this case, the Hansen coefficients $X_k^{-3,2}(0)$ with $k\neq 2$ vanish, and the rotational differential equation becomes equivalent to the one of the simple pendulum.
Adding tidal dissipations to the rotation model, the rotator can be captured in a spin-orbit resonance, where the probability of the capture depends on the width of each resonance and also on the tidal model (see Goldrich \& Peale, 1966).  Of course, for circular orbits,  the only possible final state is the synchronous rotation.   

If the circular orbital motion is now perturbed by an additional companion body, the possible rotational final states can be  very different from the Keplerian case, as it was previously shown by  Correia \& Robutel (2013).

\begin{figure}[h]
\begin{center}
 \includegraphics[width=14cm]{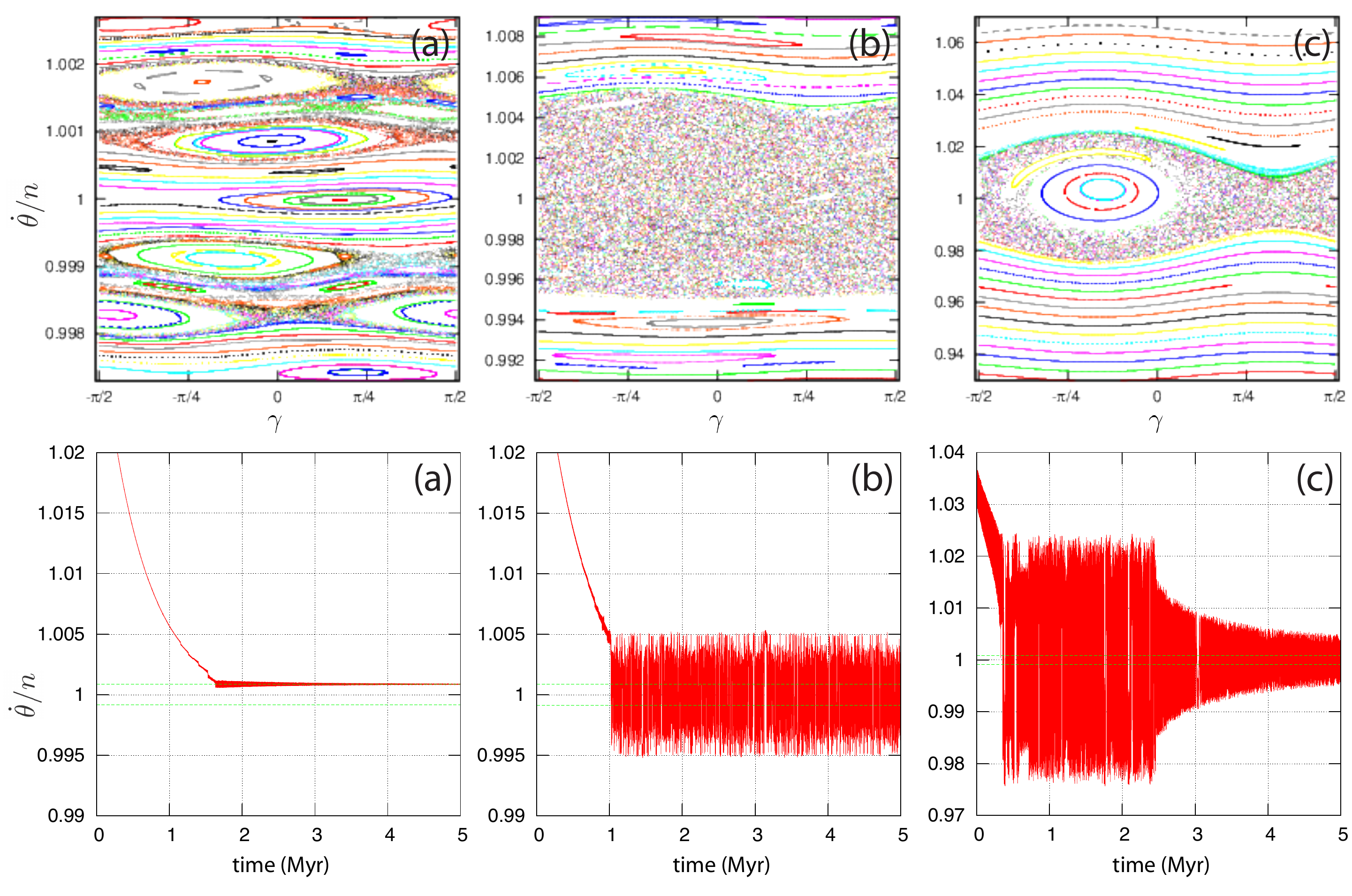} 
 \caption{
 Top: Poincar\'e surface of section in the plane of coordinates $(\gamma,\dot\theta/n)$ for $\beta = 10^{-0.5}$ (a), $10^{0.2}$ (b) and $10^{1.2}$ (c).  Bottom:  examples of tidal evolution of the rotation rate versus time. %
 \label{fig:sections_HS}  }
\end{center}
\end{figure}

\section{The coorbital case}

Let us denote $m_0$ the mass of the central body, $m_1$ the mass of the rotating body and $m_2$ the mass of its perturbing coorbital companion.
Adapting the theory developed by \'Erdi (1977) for the restricted 3-body problem to the case of three massive bodies, and neglecting the quantities of second order and more in the small quantity $\mu=(m_1+m_2)/(m_0+m_1+m_2)$, the mean longitudes and the semi-major axes of the coorbitals can be approximated by the expressions (see Robutel et al., 2011): 
\begin{equation}
 \begin{array}{ll}
\lambda_1(t) \approx   nt + \delta \zeta(t) + \lambda^{0}, & a_1(t) \approx   \bar a \left(1 -\delta \frac23 \frac{\dot\zeta(t)}{n} \right) \\
\lambda_2(t)\approx  nt - (1-\delta) \zeta(t) + \lambda^{0}, & a_2(t) \approx   \bar a \left(1 + (1-\delta)\frac23 \frac{\dot\zeta(t)}{n} \right) \\
\end{array}
\end{equation}
where $\lambda^{0}$ is a constant, $\delta = m_2/(m_1 + m_2)$, the mean semi-major axis $\bar a $ is a constant equal to $ (1-\delta)a_1 + \delta a_2$ at the fist order in $\mu$, and the variable $\zeta = \lambda_1 - \lambda_2$ satisfies the second order differential equation: 
\begin{equation}
\ddot\zeta=-3\mu n^2\left[1-(2-2\cos\zeta)^{-3/2}\right]\sin\zeta\ .
\label{eq:sol_orb_zeta} 
\end{equation}%
This differential equation, which is one of the most common representation of the coorbital motion (see Morais 1999 and references therein), possesses three equilibrium points at $(\zeta,\dot\zeta)$ equal to $(\pi/3,0)$,  $(-\pi/3,0)$ and $(\pi,0)$, corresponding respectively to the two Lagrangian equilateral configurations and to the Eulerian collinear configuration of the type $L_3$.   The separatrices emanating from this last unstable point split the phase space in three different regions: two corresponding to the tadpole trajectories surrounding one of the two Lagrange's equilibria, and another one corresponding to the horseshoe orbits which surround the three above-mentioned fixed points. As the solutions of (\ref{eq:sol_orb_zeta}) are periodic functions of frequency $\nu$ (although this frequency depends on the considered trajectory, it is always of order $\sqrt{\mu}$ and consequently small with respect to $n$),  we have: 
\begin{equation}
\left(\frac{a_1}{r_1}\right)^3\mathrm{e}^{-\mathrm{i}2\delta\zeta}=\sum_{k\in\mathbb Z}\rho_k\mathrm{e}^{\mathrm{i}(k\nu t+\phi_k)} \ ,
\end{equation}
where the amplitudes $\rho_k$ and the phases $\phi_k$ depend on the chosen trajectory. 
As in the Keplerian case, the term $\left(a/r \right)^3\sin2(\theta-f)$ appearing in (\ref{eq:rot_gene}) can be approximated at the first order in $\mu$ by the expression:
\begin{equation}
a_1^3 r_1^{-3} {\rm Imag}\left[  \mathrm{e}^{2i(\theta-f)}  \right] = a_1^3 r_1^{-3} {\rm Imag}\left[  \mathrm{e}^{2i(\theta-nt - \lambda^0)}    \mathrm{e}^{-2i\delta\zeta}\right]\, ,
\end{equation}
the rotational equation (\ref{eq:rot_gene}) can be expended in Fourier's series in the form:
\begin{equation}
\ddot\gamma=-\frac{\sigma^2}{2}\sum_{k\in\mathbb Z}\rho_k\sin(2\gamma+k\nu t+\phi_k)\ , \label{eq:rot_coorb}
\end{equation}
where $\gamma = \theta-nt - \lambda^0$ represents the orientation of the rotating body with respect to a reference frame rotating at the angular velocity $n$. 
It turns out that the main resonances can be found at $\dot\theta=n + k\nu/2$, with $k \in \mathbb{Z}$. The respective width of each associated resonant island is equal to $\sigma\sqrt{\rho _k}$.  
At one of the two Lagrange's equilibria ($\zeta=\pm\pi/3, \dot\zeta=0$) the motion of the two coorbital bodies is circular, then $a_1 = r_1 =\bar a$ and the coefficients of the expansion (\ref{eq:rot_coorb}) verify $\rho_0 = 1$ and $\rho_k = 0$ for $k\neq 0$. In this case, the equation (\ref{eq:rot_coorb}) becomes equivalent to the pendulum's equation, and as for a circular Keplerian orbit, the only possible spin-orbit resonance is the synchronous one. For small orbital librations around $L_4$ (reps. $L_5$) the  $\rho_k$ with $k\neq 0$ become strictly positive and the sequence of these coefficients, which satisfies the relations $\rho_0 > \rho_{\pm1} > \cdots > \rho_{\pm k} > \cdots $, is rapidly decreasing.  This phenomenon gives rise to new resonant islands in the both sides of the synchronous spin-orbit resonance. The widest islands are  located at  $\dot \theta = n +\nu/2$ and  $\dot \theta = n - \nu/2$, which corresponds respectively to a super-synchronous and to a sub-synchronous rotation.  The distance between these resonances being small, of the order of $\nu$ (that is, of order $\sqrt{\mu}$),  an increase in the orbital libration amplitude causes an overlapping of these resonances which can lead to chaotic rotations. 
We show in Correia \& Robutel (2013) that the global dynamics of the problem is mainly controlled by two parameters: the amplitude of the orbit around the fixed points and the quantity $\beta = \sigma/(\sqrt{\mu}n)$ which is proportional to the frequency ration $\sigma/\nu$. 

For large tadpole orbits and more specifically  for horseshoe orbits, the situation becomes more complicated than the one described above.    First of all, the sequence of the coefficients $\rho_k$ does not decrease rapidly. It follows that the width of the islands  centered at $\dot \theta = n \pm k\nu/2$ with moderate  $k$ are comparable,  which greatly enriches the rotational dynamics. 
In addition, $\rho_0$ is not necessarily the dominant coefficient and the synchronous island is no longer the widest one, as it is shown in Fig.\,\ref{fig:sections_HS}-a (top). 

The top panel  of  Fig.\,\ref{fig:sections_HS} represents the Poincar\'e sections at the time $t=2\pi/\nu$ (the coorbital period) of the differential equation (\ref{eq:rot_gene}) for three different values of $\beta$. In these three cases, the bodies whose masses verify the relation $m_1 = m_2= 10^{-6} m_0$ are on a horseshoe orbit of amplitude  ${\rm Max}\vert\lambda_1 -\lambda_2\vert=320^\circ$. This figure is equivalent to those in Correia \& Robutel (2013), although in previous work we only show the phase space for tadpole orbits.  
In Fig.\,\ref{fig:sections_HS}-a, where $\beta = 10^{-0.5}$, the phase space is dominated by $5$ islands approximatively centered at $\dot\theta = n, n \pm \nu/2, n \pm \nu$. As mentioned above, the synchronous island is not the widest one. The islands are isolated and consequently, the dynamics is globally regular. For a larger values of $\beta$ (typically of order $1$)  the main islands begin to overlap,  generating a broad chaotic zone around the synchronous rotation (Fig.\,\ref{fig:sections_HS}-b). Finally, when $\beta \gg 1$ (Fig.\,\ref{fig:sections_HS}-c),  the proximity of the main resonant islands leads to their complete overlapping. The dynamical structure is comparable to that of a modulated pendulum (see Morbidelli 2002) where the single island, which corresponds to the synchronous rotation, is surrounded by a small chaotic layer.

\section{Discussion }
By studding the rotation of a body in a quasi-circular orbit perturbed by a coorbital,  we highlight a new mechanism generating spin-orbit resonances, where the role played by the eccentricity in the case of Keplerian orbits is replaced by the amplitude of libration around the coorbital equilibria.  For a zero amplitude (Lagrange's configurations), the only possible spin-orbit resonance is the synchronization. 
The increase of this amplitude results in new resonant islands, associated to super-synchronous and sub-synchronous rotation, whose width becomes maximal in the horseshoe region.  If tidal dissipation is added to the equations (\ref{eq:rot_gene}), various final states can be reached.
In the bottom panel of the figure\,\ref{fig:sections_HS} we show examples of the rotational final states for coorbital bodies using the viscous linear model (see Mignard, 1979), whose contribution to the equation of the rotation is given by:
$
\ddot\gamma=-Ka_1^6r_1^{-6}(\dot\gamma-\delta\dot\zeta)
$, with $K=250$~yr$^{-1}$.  In these three simulations, the rotation rate that begins slightly above the synchronization ($\dot\theta/n = 1.033$), decreases before reaching its final state in a few million years.  In the case (a), the rotation ends in a super-synchronous rotation of rate $\dot\theta = n + \nu/2$. In (b), the rotation enters the large chaotic region generated by the marginal overlapping of the main spin-orbit islands, while in (c), the rotation reaches the synchronous island which is the only significant resonance for late values of $\beta$.

\end{document}